\documentclass[aps,prc,twocolumn,superscriptaddress,showpacs,10pt,dvipsnames]{revtex4-1}

\usepackage[utf8]{inputenc}
\usepackage{amsfonts}
\usepackage{amsmath}
\usepackage{amssymb}
\usepackage{cancel}
\usepackage{bm}
\usepackage[pdftex]{graphicx}
\usepackage[dvipsnames]{xcolor}
\usepackage[normalem]{ulem}
\usepackage{comment}

\begin{document}


\title{Decay of three-body resonances in a discrete basis} 




\author{J. Casal}
\email{jcasal@us.es}
\affiliation{Departamento de F\'{\i}sica At\'omica, Molecular y Nuclear, Facultad de F\'{\i}sica, Universidad de Sevilla, Apartado 1065, E-41080 Sevilla, Spain} 
\author{J. G\'omez-Camacho}
\email{gomez@us.es}
\affiliation{Departamento de F\'{\i}sica At\'omica, Molecular y Nuclear, Facultad de F\'{\i}sica, Universidad de Sevilla, Apartado 1065, E-41080 Sevilla, Spain} 
\affiliation{Centro Nacional de Aceleradores, U.~Sevilla, J.~Andalucía, CSIC, Tomas A.~Edison 7, E-41092 Sevilla, Spain} 


\date{\today}


\begin{abstract}
    We present a theoretical framework for calculating the asymptotic properties and decay dynamics of three-body resonances described in a discrete basis. The method involves solving an inhomogeneous Schrödinger equation to determine the non-normalizable resonant state by identifying a normalizable source state, which captures the short-range internal structure. The long-range behavior is then calculated using the free three-body propagator, providing accurate asymptotic coefficients necessary for describing decay correlations. We apply this formalism to the two-neutron decay of the 0$^{+}$ ground-state and the 2$^{+}$ excited-state resonances of $^{16}\text{Be}$ ($^{14}\text{Be}+n+n$), working within the hyperspherical expansion method with an analytical transformed harmonic oscillator basis. Our results show that the decay is strongly dominated by the lowest hypermomentum components at large separations, reflecting effective three-body barrier penetration dynamics that shape the final state. The calculated relative-energy distributions exhibit clear neutron-neutron correlations for both states, arising from mixing between different asymptotic channels, and are consistent with a direct two-neutron emission mechanism, in agreement with recent experimental observations. This work provides a reliable tool for linking the internal structure of three-body resonances to their decay properties.
\end{abstract}


\maketitle


\section{Introduction}

The exploration of many-body resonances in atomic nuclei poses a remarkable challenge, both experimentally and theoretically. On the experimental side, states beyond particle thresholds are often extremely short‐lived, embedded in the continuum, and must be reconstructed via invariant‐mass or complementary techniques under conditions of low production yields. On the theoretical side, the description of the states needs to address complications arising from few- or many-body asymptotics, initial nucleon-nucleon correlations and final-state interactions. A proper theoretical interpretation of experimental data typically involves an understanding on the cross-section excess around resonance energies and the dynamics driving the specific shape of relative energy and/or momentum distributions between the emitted particles.

In particular, very neutron-rich systems near and beyond the dripline provide a valuable testing ground for the study of correlated multi-neutron decays. The case of two-neutron emission has attracted considerable interest recently, as it is not yet clear how initial-state correlations are translated into final-state observables. When the single-neutron separation energy is positive but the two‐neutron separation energy is negative, the decay is expected to proceed directly, as opposed to a sequential path. It is worth noting the analogy with the structure of Borromean two-neutron halo nuclei, which are bound even though their binary subsystems are unbound. The difference here is that the nuclear interaction is not able to sustain a bound ground state. Under such conditions, the states can thus decay through the emission of two neutrons in a correlated way. A paradigmatic example is $^{16}\text{Be}~({^{14}}\text{Be} + n + n)$, which was found to be unbound with respect to the two-neutron emission threshold while lying marginally below the one-neutron threshold. This first observation of was reported by Spyrou et al.~\cite{spyrou12}, who interpreted the corresponding relative energy and angular distributions between the emitted neutrons as being consistent with a ``dineutron''-like decay. More recently, two low-lying resonances were resolved in $^{16}$Be, both decaying by direct emission~\cite{Monteagudo2024}. Another example is that of $^{26}$O, which was observed to be barely unbound with a measurable half-life and suggested the existence of two-neutron radioactivity~\cite{kohley13}, akin to the better known phenomenon of two-proton radioactivity~\cite{giovinazzo02,grigorenko09}.

Various theoretical approaches have been used to tackle the dynamics of these exotic decays. In a simple final-state interaction (FSI) model, dominated by the $nn$ scattering in the $1S_0$ channel, it was shown that the correlation function that modulates phase-space decay is sensitive to the size of the neutron source, i.e., the spatial correlations of the state~\cite{lednicky1982}. This approach was applied to study the three-body decay after Coulomb dissociation of halo nuclei~\cite{marques2001}. More sophisticated models have been developed, highlighting the effect of the structure model ingredients and dynamic approximations on the decay observables~\cite{grigorenko13,grigorenko18}. Focused on the nature of the resonant ground state, hyperspherical three-body approaches have been used to construct realistic $\text{core}+n+n$ models to extract resonance parameters and density distributions~\cite{SWang17,lovell17,JCasal18}, notably applied to $^{16}$Be and/or $^{26}$O. In Ref.~\cite{JCasal19}, we combined a pseudostate description with a so-called resonance-operator formalism, also within the hyperspherical framework, to isolate and characterize three-body resonances in a discrete basis, predicting resonance properties not only for the ground-state resonance of $^{16}$Be but also for its first 2$^+$ excited state, and focusing on the role played by dineutron correlations. Recently, a time-dependent formalism explored the evolution of the decaying resonance wave function, providing a direct way to connect initial correlations to asymptotic fragment observables~\cite{SWang21}, and confronting two-neutron and two-proton decay. It was shown that the time evolution increases the weight of the $s$-wave relative component, which is a genuine characteristic of these three-body decays, with the correlated $NN$ pairs involving mixing of different angular momenta.

In this work we present a formalism to describe the correlations between the fragments in the decay of three-body resonances, which puts on a firm ground previous works presented in Refs.~\cite{JCasal19,Monteagudo2024}. In section II we introduce a general formalism, where a broad concept of resonance is employed, applicable when an arbitrary collision of nuclei produces a set of particles $S$ in which we focus, along with other set of particles $X$. The resonance is associated to an excess in the production of particles $S$, which we will describe as the {\em system}, while the rest of the degrees of freedom constitute the {\em environment}. 
The evolution of the scattering states of the system is governed by an inhomogeneous equation, which involves the full Hamiltonian of the system plus a source term where the environment variables are integrated over. The source term is described in terms of a square normalizable state belonging to the Hilbert space of the system, which we name as the {\em source state}. The inhomogeneous equation can be rewritten in terms of the free Hamiltonian of the system, with a modified source term which involves the interactions between the particles of the system. 
To that end, the source state, which is not an stationary state of the system Hamiltonian, is constructed under the condition that it maximizes the modified source term. 
This leads to the definition of a resonance operator in the system Hilbert space, the eigenstates of which can be used as source states. These can be efficiently obtained by making use of a discrete basis for the three-body system. 
In section III we work out the expressions of the asymptotic coefficients describing the resonance states. They are determined by the coefficient of the modified source states, modulated by the properties of the Bessel functions that describe the dynamics of the free three-body system in the hyperspherical coordinate basis. Once the asymptotic coefficients are known, correlations between the observed three-body variables can be obtained. In section IV the formalism is applied to the unbound nucleus $^{16}$Be ($^{14}\text{Be}+n+n$), describing the energy correlations between the fragments observed in the 0$^+$ ground-state and 2$^+$ excited-state resonances.

\section{Resonance formalism}

We consider a general nuclear process $A+B \to S + X$, where two interacting bound systems $A$ and $B$ produce a resonant state $S$ and a residual set of particles $X$. The resonance picture is associated with an excess of cross section when the energy of the nucleons forming $S$ is in the vicinity of  a certain value $\varepsilon_r$. When the particles constituting $S$ are sufficiently far apart, they may be described by a complex-energy solution of the $S$ Hamiltonian characterized by a complex energy, $E_r = \varepsilon_r - i \Gamma/2 $ \cite{MichelGSM,taylor2006}. This solution would be a non-stationary and non-normalizable  state $|\psi^s_r \rangle$. Experimentally, the constituents of $S$ would be measured at large separations, in its center-of-mass system, and their energy distribution would be given approximatel by a Breit-Wigner shape with an energy centroid at $\varepsilon_r$ and a width $\Gamma$. 
The resonant state decays following the exponential factor $e^{- \Gamma t/ 2 \hbar }$. This, however, is compensated by its complex momentum $k = k_r - i k_i$, giving rise to a wavefunction which increases exponentially with the separation $\rho$ as $e^{k_i \rho }$. When $\rho \simeq v t$, being $v = \frac{\hbar k_r}{\mu}$ the classical velocity of separation of the fragments, both factors balance out. Note that $\hbar^2 k^2 / 2 \mu = E_r $, where $\mu$ is the relevant reduced mass, while $\rho$ is the coordinate describing the separation of all the fragments in the system $S$. If $S$ is composed of three or more fragments, $\rho$ is the hyper-radius~\cite{Zhukov93,Nielsen01}.

\subsection{Inhomogeneous equation}\label{sec:inhom}

Using the language of open quantum systems ~\cite{Okolowicz2003,MichelGSM}, the degrees of freedom associated to the particles in $S$ may be identified as the ``system'', while the degrees of freedom associated to the particles in the residual system $X$, as well as the $S$-$X$ relative coordinates, can be named as the ``environment''.  The complete reaction degrees of freedom correspond to the ``universe''. 
The states of the environment will be denoted by $|\chi^e \rangle $. Indeed, all the states of the system (resonant or non-resonant) and those of the environment could be expanded in terms of suitable biorthogonal bases of the corresponding Hilbert spaces, which allow to define \textit{bras} for the \textit{kets} introduced here~\cite{berggren68,moiseyev2011}.

The state of the universe (system + environment) is denoted by $| \Psi_T \rangle \rangle$, where the double angle is used to indicate that the Hilbert space for the universe is the product of the Hilbert spaces of both the system and the environment. 
This state will be an eigenstate of the total Hamiltonian $H_T$ corresponding to a positive, real, total energy $E_T$. Its boundary conditions are a plane wave in the partition of the projectile and target, $A+B$, with a given initial momentum $\boldsymbol{k}_{in}$, plus the outgoing waves in all the possible partitions, among them $S+X$. 

The complete state can be written in terms of a resonant component ($r$) and a non-resonant component ($nr$):
\begin{equation}
  | \Psi_T \rangle \rangle =  |\psi^s_r \rangle \times  |\chi^e_r \rangle +  |\Psi_{nr} \rangle \rangle ,  
\end{equation}
where the environment state $ |\chi^e_r \rangle $  associated to the resonance is given by
\begin{equation} 
|\chi^e_r \rangle = \langle  \psi^s_r  | \Psi_T \rangle \rangle.
\end{equation}
This bracket notation indicates integration with respect the system degrees of freedom, so the result is a state in the environment Hilbert space.  The non-resonant component can be formally expanded into non-resonant states 
$|\psi^s_{nr} \rangle$, orthogonal to the resonant state, 
\begin{equation}  
| \Psi_{nr} \rangle \rangle =  \sum_{nr} |\psi^s_{nr} \rangle \times  |\chi^e_{nr} \rangle.
\end{equation}
Although the environment states  $ |\chi^e_r \rangle,  |\chi^e_{nr} \rangle  $ may not be orthogonal, a biorthogonal basis can be found, containing a 
state $ \langle \tilde \chi^e_r | $ that fulfills 
\begin{eqnarray}
   \langle \tilde \chi^e_r  | \chi^e_{r}  \rangle &=&  1, \\
    \langle \tilde \chi^e_r  | \chi^e_{nr}  \rangle &=&  0,
\end{eqnarray} 
where the notation $ \langle \chi_i^e | \chi^e_j \rangle $ indicates partial integration over the environment variables. This leads to
\begin{equation}
    \langle \tilde \chi^e_r  | \Psi_{T} \rangle \rangle =  |\psi^s_r \rangle .
\end{equation} 

The complete Hamiltonian $H_T$ can be formally splitted into $H$, depending only on the system degrees of freedom, and $H_e$, which contains both environment and system degrees of freedom. The environment Hamiltonian $H_e$ will be such that, when the constituents of the system $S$ are far apart (i.e., when $\rho$ is very large), $H_e$ is independent on the system degrees of freedom. This excludes situations in which fragments of the system end up bound to fragments of the environment.    The Schrodinger equation becomes
   \begin{equation}
   \begin{split}
        & (H_T - E_T)  | \Psi_{T} \rangle \rangle \\  & =(H-E_r) | \Psi_{T} \rangle \rangle + (H_e - (E_T -E_r)) | \Psi_{T} \rangle \rangle = 0.
    \end{split}
   \end{equation}
Then, we can project on $\langle \tilde \chi^e_r |$, which depends only on the environment degrees of freedom, and so commutes with $(H-E_r)$. Hence, we get
\begin{equation}
   (H-E_r)  |\psi^s_r \rangle  = -  \langle \tilde \chi^e_r  |  (H_e - (E_T -E_r)) | \Psi_{T} \rangle \rangle   \end{equation}
Note that the term in the right hand side is a state which corresponds to the Hilbert space of the system, as the environment degrees of freedom are integrated. Thus, we can write
\begin{equation}   -  \langle \tilde \chi^e_r  |  (H_e - (E_T -E_r)) | \Psi_{T} \rangle \rangle  = \lambda |\phi^s_r \rangle .  \end{equation}
Here, $\lambda$ is a strength factor, while $|\phi^s_r \rangle  $ is a certain normalizable state in the Hilbert space of the system, which will be later determined. 
We consider that, when the separation $\rho$ becomes large, the wavefunctions associated to $|\phi^s_r \rangle  $, which  are  $ \langle \beta, \rho |\phi^s_r \rangle $, decrease sufficiently fast. Here, $\beta$ indicates other discrete indexes necessary to specify the state, such as angular (or hyperangular) momenta. 
Note that, as mentioned before, $H_e$ becomes independent on the system degrees of freedom when $\rho$ is large. 
In that case, the environment degrees of freedom uncouple from those of the system, and $\langle \tilde \chi^e_r  |  $ becomes an eigenstate of $H_e$ associated to the available energy $E_T-E_r$. Thus, $\langle \tilde \chi^e_r  |  (H_e - (E_T -E_r)) \simeq 0$ when $\rho$ is large, and so $ \langle \beta, \rho |\phi^s_r \rangle $ vanish.



Hence, the equation that describes the system displaying the resonant behavior is:
  \begin{equation}
     (H-E_r)  |\psi^s_r \rangle  = \lambda |\phi^s_r \rangle.  
  \end{equation} 
The resonant state, $ |\psi^s_r \rangle $ understood as a distribution of continuum states producing an increase in the cross section, is defined in terms of the system Hamiltonian $H$, but also in terms of a certain square normalizable state appearing as a source term in the Schrödinger equation. All the information of the total scattering energy $E_T$, projectile $A$, target $B$, and interactions $H_e$ between system and environment is embedded in the quantity $\lambda$. We do not pretend here to enter in the evaluation of this parameter, which, as we shall see, may affect the magnitude of the cross section, but not the correlations of the relative coordinates between the particles constituting the resonance. Note that different reactions, with different choices of projectile,  target or scattering energies $E_T$, may influence the magnitude of $\lambda$. Moreover, if the environment variables are somehow measured (for example, by measuring the relative momentum of $S$ and $X$), the value of  $\lambda$ will depend on these measurements, and so will the overall cross sections for the resonance production. 
Since we will focus on the description of the system, ant not the environment, we may drop the superindex $s$ and focus on the solutions of the equation
\begin{equation}
    (H-E_r)  |\psi_r \rangle  = \lambda |\phi_r \rangle .  
    \label{eq:inhom1}
\end{equation} 

\subsection{Definition of the source state}
In Eq.~(\ref{eq:inhom1}), $|\psi_r \rangle$ represents the long-range, non-normalizable resonant state, which will asymptotically describe the correlations among the fragments of the system in its decay, and $E_r$ is the corresponding complex energy. 
In the right-hand side, we will refer to $ |\phi_r \rangle $ as the source state, which is normalizable according to the discussion above. 

We can now distinguish the long range behavior of the resonant state $ |\psi_r \rangle  $ from its short-range behavior. For the latter, a discrete, finite, but arbitrarily large basis $|b \rangle$ can be used.  This basis can be constructed with a suitable family of orthogonal polynomials on the radial variable $\rho$ and 
the angular momentum indexes $\beta$. 
Thus, the overlaps $\langle \rho, \beta | b \rangle $ are known, and they fulfill the orthogonality condition
\begin{equation}
  \sum_\beta   \int_0^\infty  d \rho  \langle b' |\rho, \beta  \rangle \langle \rho, \beta | b \rangle = \delta_{b b'}.
\end{equation}

The projector on the short-range subspace is \[\displaystyle P = \sum_b |b \rangle \langle b|,\] with its complementary projector being $Q = 1 - P$. As we argued previously, the source state $ |\phi_r \rangle  $ is short range, so we can approximate $ |\phi_r \rangle  \simeq P |\phi_r \rangle  $. Then, we get 
\begin{eqnarray}
P (H- E_r) P |\psi_r \rangle  &+&  P (H- E_r)  Q|\psi_r \rangle  =  \lambda |\phi_r \rangle, \nonumber \\
Q (H- E_r) P |\psi_r \rangle  &+&  Q (H- E_r)  Q|\psi_r \rangle  = 0,
\end{eqnarray}
from which we obtain the equations
\begin{eqnarray}
\lambda |\phi_r \rangle  & = &  (H_{ef}- E_r) P |\psi_r \rangle \nonumber, \\
H_{ef} &=&   P H P -  P H Q \frac{1}{Q (H- E_r) Q} Q H P \label{hef}, \\
 P|\psi_r \rangle &=& \frac{\lambda}{H_{ef}- E_r}  |\phi_r \rangle.
\end{eqnarray}
Note that the effective Hamiltonian $H_{ef}$ is an operator defined within the short-range subspace, which is obtained from the complete Hamiltonian $H$ using the Feshbach formalism \cite{Feshbach}.
Although it formally depends on the matrix elements of the true, complete Hamiltonian $H$, beyond the short-range space, in practical terms the effective Hamiltonian can be approximated in terms of effective interactions restricted to the short-range subspace. 
Thus, we will assume that the effective Hamiltonian $H_{ef}$ is known, and can be constructed taking into account kinetic energy terms plus effective interactions. The eigenstates $|n \rangle$, corresponding to the real eigenvalues $\varepsilon_n$, and their overlaps $ \langle \rho, \beta | n \rangle $ can also be obtained. 
Since the projector $P$ can be written in the $| n\rangle$ basis as $P = \sum_n |n \rangle \langle n|$, and $\frac{1}{H_{ef}- E_r}$ is diagonal in that basis, we get
\begin{equation}
    \langle n |\psi_r \rangle  =   \frac{\lambda}{\varepsilon_n-E_r} \langle n|\phi_r \rangle.
\end{equation}

The system Hamiltonian can be written as $H = H_0 + V$, where $H_0$ contains the kinetic energy, and, possibly, a long-range Coulomb interaction, while $V = PVP$ is the short-range interaction. Therefore 
\begin{equation}
     (H_0-E_r)  |\psi_r \rangle  =   -V  |\psi_r \rangle  + \lambda |\phi_r \rangle . 
\end{equation}  
The right-hand side is short range, so it can be simply projected into the discrete basis. Thus, we have
\begin{equation}
   (H_0-E_r)  |\psi_r \rangle  = \lambda |\Phi_s \rangle, \label{eq:psi}
\end{equation}
   where 
\begin{equation}
   |\Phi_s \rangle  =  \sum_{n m} |n \rangle \left(\frac{ -\langle n | V | m \rangle}{\varepsilon_m -E_r} + \delta_{nm} \right) \langle m |\phi_r \rangle .   
   \label{eq:Phi}
\end{equation}
We will refer to $|\Phi_s \rangle$ as the modified source term which includes the system interactions. A compact expression of this term, in terms of the kinetic energy operator $T = H_{ef} - V$, which acts in the short range subspace, is
\begin{equation}
   |\Phi_s \rangle  =  (T- E_r) \frac{1}{H_{ef}- E_r} |\phi_r \rangle    \label{eq:Phioper}
\end{equation}
Note that the operator $T$ as well as $H_{ef}$ are restricted to the short-range subspace. They are then different from the operators $H_0$ and $H$, which act on the full space. The interaction $V$, however, is short range, so it is the same both spaces. 

As it will be mentioned later, there can be some advantages in considering an equation similar to Eq.~(\ref{eq:psi}) in which the complex energy $E_r$ is substituted by its real part. This is achieved moving the term $i \frac{\Gamma}{2} |\psi_r \rangle $ to the source term
\begin{equation}
   (H_0-\varepsilon_r)  |\psi_r \rangle  = \lambda |{\Phi}'_s \rangle, 
   \label{eq:psiereal}
\end{equation}
where
\begin{equation}
   |{\Phi}'_s \rangle  =  \sum_{n m} |n \rangle \left(\frac{ -\langle n | V | m \rangle - i {\Gamma \over 2} \delta_{nm}}{\varepsilon_m -E_r} + \delta_{nm} \right) \langle m |\phi_r \rangle .
   \label{eq:Phitilde}
\end{equation}
We will refer to $|{\Phi}'_s \rangle$ as the modified source term which includes the system interactions and the resonance width. A compact expression is
\begin{equation}
   |{\Phi}'_s \rangle  =  (T- \varepsilon_r) \frac{1}{H_{ef}- E_r} |\phi_r \rangle    \label{eqPhioper}
\end{equation}
Note that we have a term $(T-\varepsilon_r)$, where $\varepsilon_r$ is the real part of the resonance energy $E_r$.
 
Let us reflect on the meaning of these equations. The left hand side of Eqs.~(\ref{eq:psi}) and (\ref{eq:psiereal}) contain the operator $H_0$, as well as the state  $ |\psi_r \rangle  $, which describes the resonance at all distances, 
including the correlations between the fragments observed at large distances, 
in correspondence with eventual experimental observations after the production and decay of the resonance. 
In the next section, the equation will be solved taking into account that $H_0$ involves differential operators, so no $P$ projection will be done in $H_0$. 
On the contrary, $ |\Phi_s \rangle $ is a short range term, which is calculated in the finite basis. 
$ |{\Phi}'_s \rangle $ is also short range, provided that the term $i\frac{\Gamma}{2}|\psi_r\rangle $ is approximated by  $i\frac{\Gamma}{2}P |\psi_r\rangle $, which is consistent with the fact that the value of $\Gamma$ may be derived in a finite basis. Other authors also use this assumption when formulating inhomogeneous equations in this context~\cite{grigorenko13,grigorenko18}. 
Note that $ |\Phi_s \rangle$, $|{\Phi}'_s \rangle $ have the meaning of  source terms for the corresponding differential equations. 
The requirement to evaluate $ |\Phi_s \rangle$ (or $|{\Phi}'_s \rangle $) is to determine the source state $|\phi_r \rangle$, which is a normalizable state, described in the discrete basis. 
Once the source state $|\phi_r \rangle$ is known, we can evaluate the source terms 
and thus, apart from a normalization factor $\lambda$, the resonance state $|\psi_r \rangle$ can be determined at arbitrarily large distances.

The question that remains is how to determine the source state. In the context of the production and decay of a resonant state in a given nuclear reaction, the resonance is typically linked to an increase in the cross section. 
Within our formalism, a significant increase in the cross section for the resonance production will occur provided that the source term is large. Thus, we will focus on source states $|\phi_r \rangle$ that maximize the norm of the source term, $\langle \Phi_s |\Phi_s \rangle$ (or $\langle {\Phi}'_s |{\Phi}'_s \rangle $). 
Note that alternative choices for the source term describing a resonance have been proposed in the literature.  In Refs.~\cite{grigorenko13,grigorenko09}, the inhomogeneous equation involves a source term given by a solution of the effective Hamiltonian with box boundary conditions at the precise (real) resonance energy. We do not claim that our choice for the source term is guaranteed to give the best description for the resonance wave function at large distances.  However, our approach provides a robust procedure to identify resonances from the short range description based on an arbitrary square normalizable basis of pseudostates, and to extrapolate its behavior to long distances.

The source states that give maximum norm of the source term can be formally obtained as eigenstates of specific operators, as shown in the Appendix. 
Provided that the energy and width of the resonance is small, these operators are closely related to the resonance operator $M = H_{ef}^{-1/2} V  H_{ef}^{-1/2}$, proposed heuristically in Ref.~\cite{JCasal19}. The eigenstates of this operator are then identified with the source states of the resonance. The time evolution of those non-stationary states in the pseudostate basis, as shown in Ref.~\cite{JCasal19}, determines the complex energy $E_r=\varepsilon_r-i\frac{\Gamma}{2}$ of the resonance.

\section{Three-body Resonance asymptotics}

To obtain the resonance we need to solve Eq.~(\ref{eq:psi}). However, it is not sufficient to solve it in the short range subspace defined by the projector $P$, because we want to get the long range behavior required to get the asymptotic behavior. The source term in Eq.~(\ref{eq:psi}), however, is short range, and thus it can be taken in the subspace. Formally, this can be done making use of the free propagator $G_0(E_r)  = (H_0-E_r)^{-1} $
\begin{equation}
  |\psi_r \rangle  = \lambda G_0(E_r)|\Phi_s \rangle  \label{eqpro}
\end{equation}
We can  expand the states $| \psi_r \rangle$ in terms of the radial channel basis $|\beta, \rho \rangle$ introduced in Sec.~\ref{sec:inhom}. For a three-body system, this can be done using the hyperspherical formalism, which we outline in the next section. 

\subsection{Three-body hyperspherical formalism}

Within the hyperspherical formalism, normalizable three-body states for a given total angular momentum $j$ can be expanded in an arbitrarily large basis $|b,j\rangle$ with overlaps
\begin{equation}
\langle\beta,\rho,\Omega|b,j\rangle=\rho^{-5/2}U_{b\beta}(\rho)\mathcal{Y}_{\beta}^{j\mu}(\Omega).
\label{eq:basisbeta}
\end{equation}
Here, $\rho=\sqrt{x^2+y^2}$ is the hyper-radius given in terms of usual Jacobi coordinates $\left\{\boldsymbol{x},\boldsymbol{y}\right\}$, and $\Omega\equiv\{\alpha,\widehat{x},\widehat{y}\}$, with $\alpha$ the hyper-angle given by  $\tan\alpha=y/x$. Note that there are some mass factors that relate the scaled Jacobi coordinates with actual physical distances~\cite{Zhukov93}. The channel index $\beta$ contains all quantum numbers associated to the angular momentum couplings compatible with $j$, and $\mathcal{Y}_{\beta}^{j\mu}$ is consistently given in terms of hyperspherical harmonics $\Upsilon_{Klm}^{l_xl_y}$. A common choice for the coupling order is
\begin{equation}
\mathcal{Y}_{\beta}^{j\mu}(\Omega)=\left\{\left[\Upsilon_{Kl}^{l_xl_y}(\Omega)\otimes\phi_{S_x}\right]_J\otimes\kappa_I\right\}_{j\mu},
\label{eq:Upsilon}
\end{equation}
where $\boldsymbol{l}=\boldsymbol{l}_x+\boldsymbol{l}_y$ is the total orbital angular momentum, $S_x$ is related to the coupled spin of the two particles in the $x$ coordinate, $\boldsymbol{J}=\boldsymbol{l}+\boldsymbol{S}_x$, and $I$ labels the spin of the third particle. The hyperspherical harmonics, eigenstates of the so-called hypermomentum operator $\widehat{K}$, are
\begin{equation}
\Upsilon_{Klm}^{l_xl_y}(\Omega)=\varphi_K^{l_xl_y}(\alpha)\left[Y_{l_x}(\widehat{x})\otimes Y_{l_y}(\widehat{y})\right]_{lm},
\label{eq:HH}
\end{equation}
\begin{equation}
\begin{split}
\varphi_K^{l_xl_y}(\alpha) & = N_{K}^{l_xl_y}\left(\sin\alpha\right)^{l_x}\left(\cos\alpha\right)^{l_y}\\&\times P_p^{l_x+\frac{1}{2},l_y+\frac{1}{2}}\left(\cos 2\alpha\right),
\label{eq:varphi}
\end{split}
\end{equation}
where $N_K^{l_xl_y}$ is a normalization constant and
$P_p^{a,b}$ represents a Jacobi polynomial of order $p=(K-l_x-l_y)/2$. It is worth noting that the hypermomentum $K$ determines the effective three-body barriers that appear in the effective Hamiltonian. 

For the hyperradial basis functions in Eq.~(\ref{eq:basisbeta}), $\langle\beta,\rho|b\rangle=U_{b\beta}(\rho)$, in this work we use the analytical Transformed Harmonic Oscillator (THO) basis~\cite{JCasal13}. In the case of eigenstates of the effective three-body Hamiltonian associated to the eigenvalues $\varepsilon_n$, we refer to pseudostates $|n\rangle$, and their radial dependence is given by a combination of the known basis functions
\begin{equation}
    \langle\beta,\rho|n\rangle=\sum_{b}\langle\beta,\rho|b\rangle\langle b|n\rangle = \sum_b C_{b\beta}^{n}U_{b\beta}(\rho).
\end{equation}
Then, the source state will be obtained as a localized eigenstate of the resonance operator discussed in the previous section, and will be expanded in the basis of energy pseudostates, i.e.,  
\begin{equation}
    |\phi_r\rangle = \sum_n\mathcal{C}_n|n\rangle.
    \label{eq:expandphi}
\end{equation}
In practice, coefficients $C_{b\beta}^{n}=\langle b|n\rangle$ (for each $\beta$) and $\mathcal{C}_n=\langle n|\phi_r\rangle$ will be obtained by diagonalizing the three-body Hamiltonian and the resonance operator, respectively. The reader is referred to Ref.~\cite{JCasal19} for specific details about the construction of the source state for the resonance. From that state, the source term in configuration space is given by the following hyperradial functions for each $\beta$ channel,
\begin{align}
    & \langle\beta,\rho|{\Phi}'_s \rangle  =  \sum_{n}  \langle\beta,\rho|n \rangle d_n,  \label{eq:sourcerho2} \\
    & d_n =\sum_m\left(\frac{ -\langle n | V | m \rangle + (\varepsilon_m -\varepsilon_r) \delta_{nm}}{\varepsilon_m -\varepsilon_r + i\frac{\Gamma}{2}} \right) \langle m |\phi_r \rangle.
    \label{eq:sourcerho}
\end{align}

\subsection{Asymptotic coefficients}

Note that the free Hamiltonian $H_0$ is diagonal in the index $\beta$, although it is not diagonal in the hyper-radius $\rho$. The propagator, relevant for outgoing boundary conditions of the resonance, can be written as
\begin{equation}
\langle \beta, \rho | G_0(E_r) | \beta, \rho' \rangle =  \frac{4 m i}{\pi \hbar^2 k} F_\beta(k \rho_<) H^+_\beta( k \rho_>).
\end{equation}
Here, $k$ is the complex momentum defined in terms of the resonance energy as $\frac{\hbar^2 k^2}{2 m} = E_r$, while
$F_\beta(x)$ and $H^+_\beta(x)$ are the regular and outgoing solutions of $(H_0-E_r) \phi_\beta(x) = 0$. Note that these solutions depend only on the index $K$ contained in $\beta$. Indeed, a Coulomb-like term proportional to $\rho^{-1}$ can be included in $H_0$, and then  $F_\beta(x), H^+_\beta(x)$ will be Coulomb wave functions. The explicit expressions of  $F_\beta(x), H^+_\beta(x)$ depend on the number of orbital degrees of freedom relevant for the resonance. 
For three-body systems without Coulomb interaction, $F_\beta(x), H^+_\beta(x)$ are given, respectively, in terms of Bessel and Hankel functions $x ^{1/2} J_{K+2}(x), x^{1/2} H^+_{K+2}(x)$.  

The radial behavior of $  |\psi_r \rangle $ in the different channels is given by
\begin{equation}
\langle \beta, \rho   |\psi_r \rangle  =  \lambda \int d \rho'  \langle \beta, \rho  | G_0(E_r)|  \beta, \rho' \rangle     \langle \beta, \rho' | \Phi_s \rangle .\label{eq:radt}
\end{equation}
As the distances $\rho$ related to the long-range correlations are much larger than the distances $\rho'$ associated to the source term, we get
\begin{eqnarray}
\langle \beta, \rho   |\psi_r \rangle  &=&  \frac{4 m i \lambda}{\pi \hbar^2 k} \sqrt{k \rho} H^+_{K+2}(k \rho) A_\beta, \\
A_\beta & = & \int d \rho'    \sqrt{k \rho'} J_{K+2}(k \rho')   \langle \beta, \rho'  |\Phi_s \rangle.  \label{eq:rad}
\end{eqnarray}
The overlaps $ \langle \beta, \rho'  |\Phi_s \rangle   $  can be obtained using Eq.~(\ref{eq:Phi}) from those of the eigenstates $\langle \rho, \beta | n \rangle$. 
Note that, as $k= k_r - i k_i$ is complex, the Hankel function $H^+_{K+2}(k \rho)$ increases exponentially as $\exp(+ k_i \rho)$. However, as it was mentioned previously, this exponential factor cancels exactly with the time dependence $\exp(- \frac{\Gamma}{2} t)$, when $\rho \simeq v t$, while the velocity is $v = \hbar k_r / m$, which is the velocity of a wave packet made of a narrow range of $k$ values. So, the use of complex momenta should not be problematic, as far as the long $\rho$ behavior is concerned. However, in the evaluation of the amplitude $A_\beta$ using Eq.~(\ref{eq:rad}), the Bessel function $J_{K+2}(k \rho) $ also grows exponentially, and this may generate numerical problems in the convergence of the integral. 
Thus, we may use the expressions obtained assuming that the energy in the propagator is purely real. This is achieved moving the imaginary part of the energy to the source term, as discussed in the previous section, leading to the different source term $|\Phi'_s \rangle$ in Eq.~(\ref{eq:Phitilde}).  Then, the equations become
\begin{eqnarray}
\langle \beta, \rho   |\psi_r \rangle  &=&  \frac{4 m i \lambda}{\pi \hbar^2 k} \sqrt{k_r \rho} H^+_{K+2}(k_r \rho) A'_\beta, \\
A'_\beta & = & \int d \rho'    \sqrt{k_r \rho'} J_{K+2}(k_r \rho')   \langle \beta, \rho'  |\Phi'_s \rangle.  \label{eq:radkr}
\end{eqnarray}

\subsection{Energy correlations}

The above expressions provide the asymptotic coefficients, which give the large-$\rho$ behavior in terms of outgoing functions. We are interested, however, in describing the relative-energy distributions between the fragments in the decay. This can be obtained from the hyperangular distribution of the asymptotic state, assuming it is the same in coordinate and in momentum space~\cite{grigorenko03,grigorenko09}, apart from normalization and phase factors. This is strictly valid for the asymptotic solution and on-shell conditions, i.e., if the solution corresponds to a precise real energy. For a resonance of complex-energy $E_r$, the approximation works as long as its energy distribution is narrow and peaked around $\varepsilon_r$. Note that the definition of the 6-dimensional plane wave in hyperspherical coordinates, involved in the Fourier transform that links coordinate and momentum spaces, includes an $i^K$ relative phase~\cite{Desc03,Zhukov93} which is required for consistency. Taking this into account, and integrating with respect to the angles $\widehat{x},\widehat{y}$, the hyperangular distribution can be written as
\begin{align}
   \nonumber \frac{d  \mathcal{P}(\alpha)}{d \alpha} & \propto \sum_{l_x,l_y, l,S_x, J, I}\sum_{K,K'} i^{K-K'} A_\beta (A_{\beta'})^*   \\
    & \times \varphi_K^{l_xl_y}(\alpha) \varphi_{K'}^{l_xl_y}(\alpha), 
    \label{eq:Palpha}
\end{align}
where $\varphi_K^{l_xl_y}(\alpha)$ are the hyperangular functions given by Eq.~(\ref{eq:varphi}) in terms of Jacobi polynomials. The relative energies in the $x$ and $y$ Jacobi subsystems are, respectively, $E_x=E\cos^2(\alpha)$ and $E_y=E\sin^2(\alpha)$, with $E=\hbar^2\kappa^2/(2m)$ and $\kappa$ the momentum conjugate of $\rho$. With the approximation that $E$ follows the narrow energy distribution of a resonance peaked at $\varepsilon_r$, from the hyperangular distribution we can obtain the relative-energy spectrum for the particles related by the $x$ coordinate as
\begin{equation}
\frac{d\mathcal{P}(\varepsilon_{x})}{d\varepsilon_{x}}=\frac{1}{\sin(2\alpha)}\frac{d\mathcal{P}(\alpha)}{d\alpha},
\label{eq:Pex}
\end{equation}
where $\varepsilon_x=E_x/\varepsilon_r$. For a decaying $\text{core}+n+n$ system, it is customary to work in the so-called Jacobi-T set where the two neutrons are related by $x$. Then, $\varepsilon_x\equiv\varepsilon_{nn}$ is the reduced neutron-neutron relative energy. Note that, using the same expression but rotating the wave functions (and asymptotic coefficients) to the Jacobi-Y set, the $\text{core}+n$ distribution ($\varepsilon_{cn}$) can be analogously obtained. Details about the wave function transformations between different Jacobi sets, which involve Raynal-Revai coefficients, can be found elsewhere~\cite{IJThompson04,Casal20,RR70}. 
Note that the distributions~(\ref{eq:Palpha},\ref{eq:Pex}) come from a coherent sum of the different asymptotic channels. This will be discussed in the next section. 

\section{Application to Beryllium-16}

After the first observation of the two-neutron decay of $^{16}$Be~\cite{spyrou12}, a more recent measurement was able to resolve two close resonances at 0.84(3) and 2.15(5) MeV above the two-neutron threshold, likely the 0$^+$ ground state and 2$^+$ first excited state, with widths of 0.32(8) and 0.95(15) MeV, respectively~\cite{Monteagudo2024}. A coincidence analysis of the fragment energy correlations was consistent with a dominant direct two-neutron decay for the two resonances. We present here the details about the three-body calculations included in Ref.~\cite{Monteagudo2024} to characterize the $nn$ decay energy distributions, focusing on the resonance asymptotic properties and expanding on those results. 

\subsection{Source states and source terms}
The $^{16}$Be ($^{14}\text{Be}+n+n$) calculations for the construction of the source state were performed as in Ref.~\cite{JCasal19}. First, pseudostates were obtained by diagonalizing the three-body Hamiltonian in a THO basis. Then, the resonances were identified by diagonalizing the resonance operator in the basis of energy pseudostates. We employed a THO basis with parameters $b=0.7$ fm and $\gamma=1.1$ fm$^{1/2}$. The convergence of the calculations was ensured by using a sufficiently large number of basis functions, and a model space fixed by $K_{max}=30$, which determines the maximum value of $l_x$ and $l_y$ in the hyperspherical harmonics expansion. The model space contained 136 $\beta$ channels for the 0$^+$ and 265 channels for the 2$^+$. Three-body coupling potentials were generated by using the $n$-$n$ tensor interaction from Ref.~\cite{GPT}, and an effective $\text{core}$-$n$ potential fixed to the known $d_{5/2}$ resonance of $^{15}$Be~\cite{lovell17}. This potential gives rise to bound states in the 1$s_{1/2}$, 1$p_{3/2}$ and 1$p_{1/2}$ orbitals, which are occupied by the core neutrons, so they were projected out of the calculations by means of a supersymmetric transformation that generates phase-equivalent potentials without Pauli states. As in Ref.~\cite{JCasal19}, an additional hyperradial three-body force was included to fix the energy of the 0$^+$ ground-state resonance at 0.85 MeV above the $2n$ threshold. Without any extra parameters, an energy of 2.15 MeV was obtained for the 2$^+$ resonance, in good agreement with the experimental separation between the two states. Note that, in Ref.~\cite{JCasal19}, we fixed the ground state to a larger energy consistent with the previous measurement~\cite{spyrou12}. In the present work, the most recent experimental results were followed. The time-evolution of these states, eigenstates of the above-mentioned resonance operator but not stationary, allowed to determine their widths as $\Gamma(0^+)=0.10$ and $\Gamma(2^+)=0.42$ MeV, both smaller than the experimental results. As discussed in Ref.~\cite{Monteagudo2024}, the underestimation of the widths might be linked to the inert-core approximation or the lack of more constrains for the $\text{core}$-$n$ potential. This remains a question for further research. 

\begin{figure}
\centering
\includegraphics[width=0.9\linewidth]{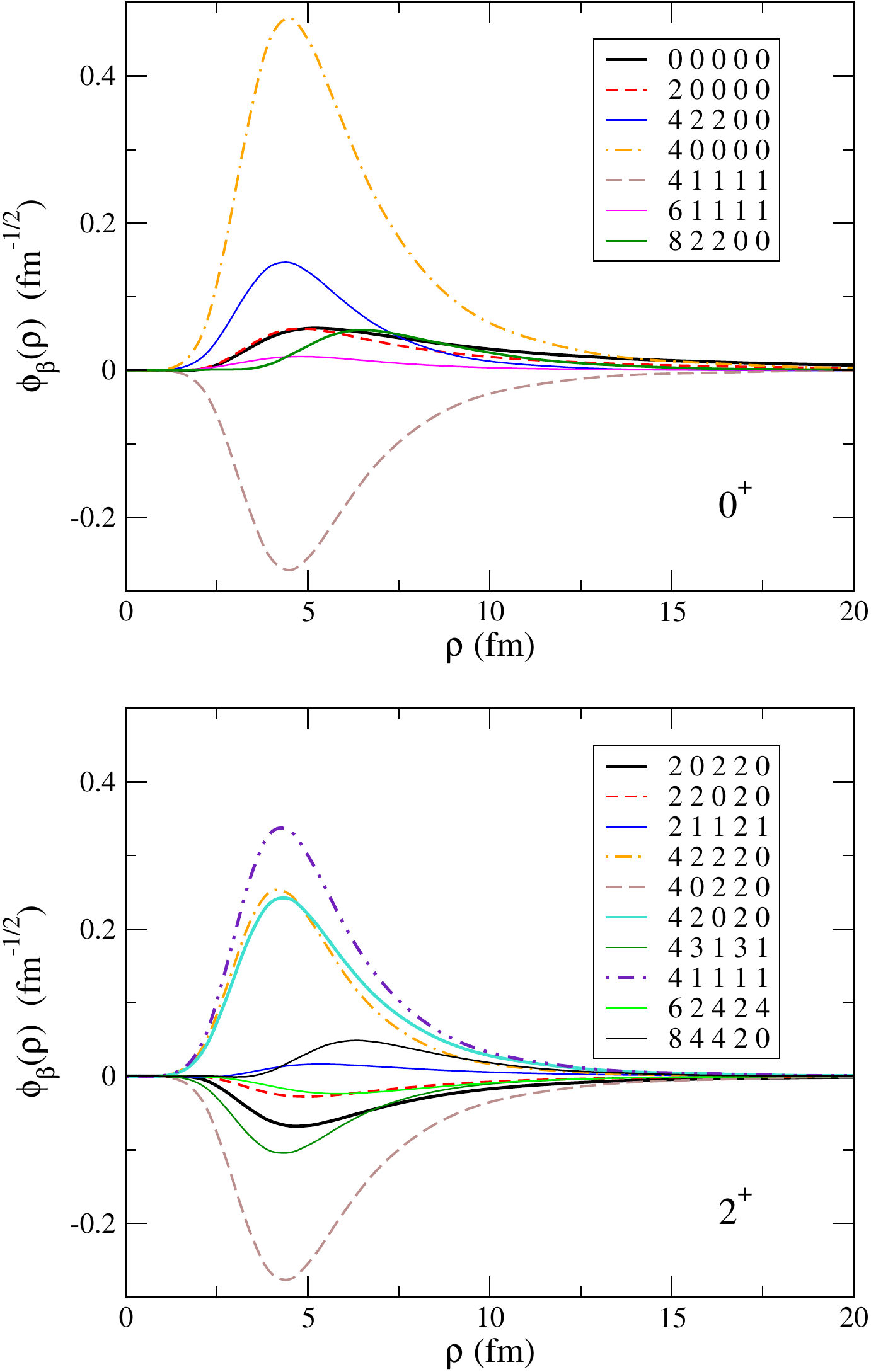}

\caption{Hyperradial wave functions of the source state for the 0$^+$ (top panel) and 2$^+$ (bottom panel) resonances of $^{16}$Be. The different lines correspond to relevant channels labeled by $\beta\equiv\{K,l_x,l_y,l,S_x\}$ in the Jacobi-T set.}
\label{fig:wfs}
\end{figure}

\begin{figure}
\centering
\includegraphics[width=0.8\linewidth]{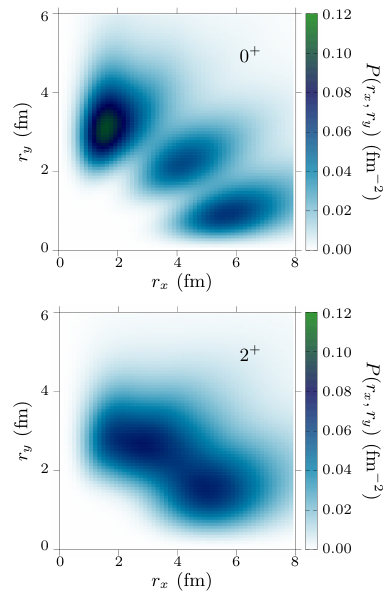}

\caption{Spatial probability density distribution of the source state for the resonances $0^+$ (top panel) and $2^+$ (bottom panel), as a function of the distances $n$-$n$ ($r_x$) and $^{14}$Be-$nn$ ($r_y$).}
\label{fig:xyprob}
\end{figure}

Figure~\ref{fig:wfs} shows the hyperradial dependence, $\phi_\beta(\rho)=\langle\beta,\rho|\phi_r\rangle$, of some relevant $\beta$ channels up to $K=8$ in the Jacobi-T representation for the $0^+$ and $2^+$ states. For $K=6$ and $K=8$, only the largest component is shown. Higher-$K$ terms were found to play a less significant role. It is clear that the $K=4$ components in the hyperspherical expansion are dominant. This is consistent with the two valence neutrons occupying mostly the $d_{5/2}$ orbital in a shell-model picture. Indeed, if the wave functions are transformed to the Jacobi-Y set where the $\text{core}+n$ subsystem is in the $x$ coordinate, the $l_x^{(\text{Y})}=2$ content is 94.8\% and 96.5\% for the 0$^+$ and $2^+$ states, respectively. In Fig.~\ref{fig:xyprob} we present also the spatial probability density of the states in the Jacobi-T system, adapted from Ref.~\cite{Monteagudo2024}. The 0$^+$ ground state presents a clear dineutron peak corresponding to small $r_x\equiv r_{nn}$ distances, while the $2^+$ state exhibits a more diffuse spatial distribution. In terms of the relative angular momentum between the two neutrons, the wave function has 70\% of $l_x=0$ for the 0$^+$ state but only 24\% for the 2$^+$. This indicates that the spatial dineutron correlations are stronger in the ground-state resonance. 

\begin{figure}
\centering
\includegraphics[width=0.9\linewidth]{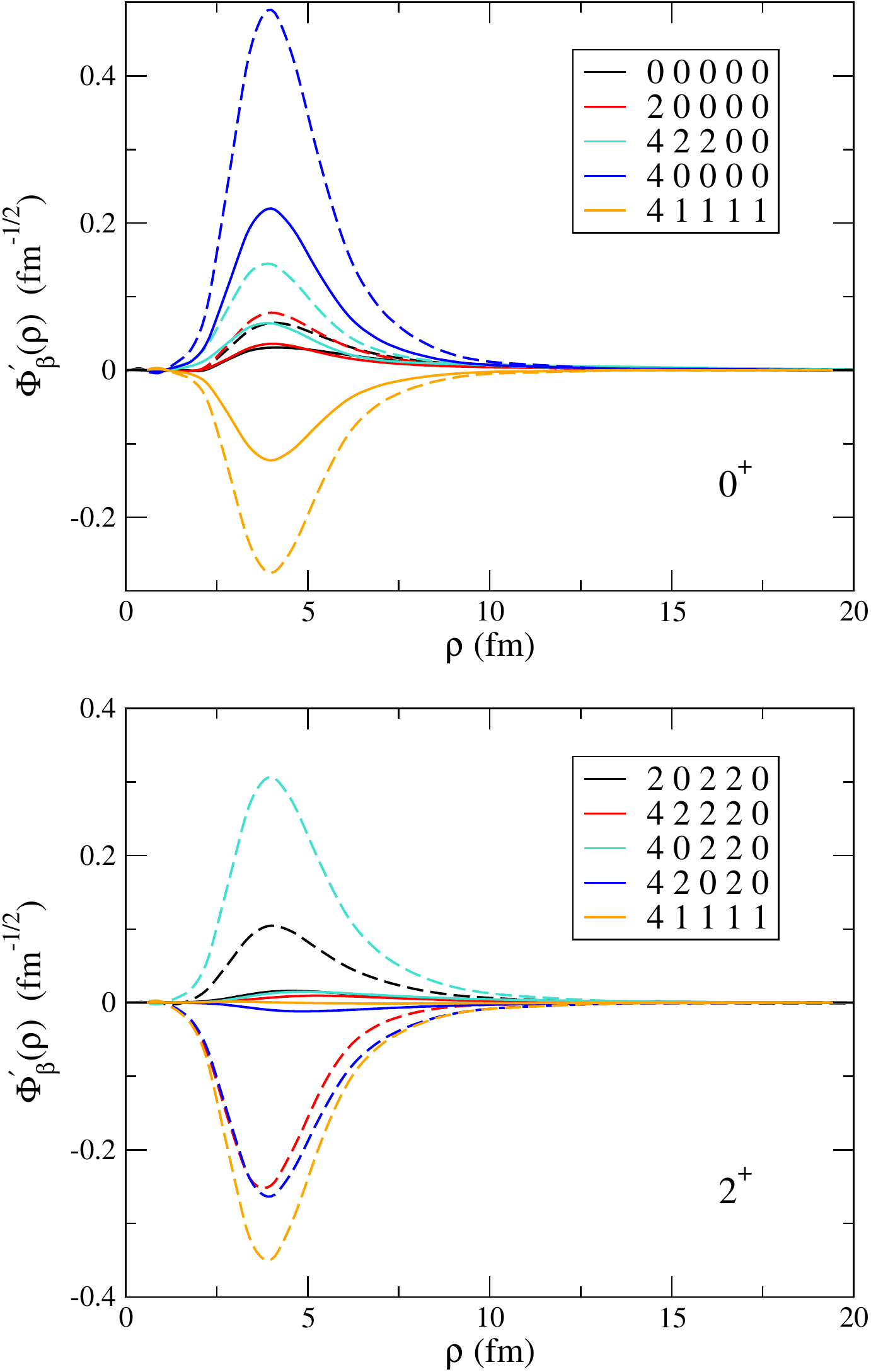}

\caption{Modified source terms for the 0$^+$ (top panel) and 2$^+$ (bottom panel) resonances. The real (solid) and imaginary parts (dashed) are shown for different channels. The labels are the same as in Fig.~\ref{fig:wfs}.}
\label{fig:source}
\end{figure}

The modified source terms are computed from the source states using Eq.~(\ref{eq:sourcerho}). Their hyperradial dependence, ${\Phi}'_\beta(\rho)=\langle\beta,\rho|{\Phi}'_s\rangle$, for the most representative channels is shown in Fig.~\ref{fig:source}. Note that they are complex functions, due to the presence of the factor $i\Gamma/2$ in the construction of the modified source term. Notice also that, in the case of the $2^+$ resonance, the imaginary parts are more significant due to the relatively larger width. The overall change of the radial dependence of the modified source terms with respect to that of the source states is a reduction of the radial extension, as can be seen by comparison with Fig.~\ref{fig:wfs}. The relative importance of the different channels in the source state is mostly preserved in the modified source term. 

From the modified source term, we can compute the asymptotic coefficients describing the resonance correlations at large distances, i.e., Eq.~(\ref{eq:radkr}). The weights of the most relevant components, normalized to unity, are shown in Table~\ref{tab:coef}. It is seen that, asymptotically, the channels with the lowest possible $K$ values in the wave-function expansion dominate. For the $0^+$ ground-state resonance, $\approx 99\%$ corresponds to the only $K=0$ channel, with a very small fraction of $K=2$ and almost negligible contribution of higher hypermomenta. In the case of the $2^+$ resonance, around $\approx 90\%$ corresponds to $K=2$ (with the channel $K=2, l_x=0$ exhausting almost 76\%),  while $K=4$ amounts to a value smaller than 8\%. Note that, due to angular momentum conservation, $K=0$ is not allowed for the $2^+$ state. These results somehow reflect the evolution from the source states, which are dominated by $K=4$ components, to asymptotic resonance wave functions that favor the lowest possible centrifugal barriers. In a sense, the formalism highlights the barrier penetration process. Note that, in the time-dependent description of Ref.~\cite{SWang21}, it was shown that the wave functions for two-nucleon decaying systems evolve by increasing the weight the lowest partial waves, which is consistent with the present results.

\begin{table}
  \centering
    \begin{tabular}{lclcll}
     \toprule
     $j^\pi$  & & $\beta$ & & $A'_\beta$ & $\left|A'_\beta\right|^2$ \\
     \colrule
     0$^+$ & &0 0 0 0 0 & & $(0.9957,0.0000)$  & 0.991\\    
           & &2 0 0 0 0 & & $(-0.0836,0.0092)$ & 0.007\\    
           & &4 2 2 0 0 & & $(0.0068,0.0292)$ & $<0.001$\\  
     2$^+$ & &2 0 2 2 0 & & $(0.8692,0.0000)$& 0.756\\    
           & &2 2 0 2 0 & & $(0.2964,-0.0695)$& 0.093\\   
           & &2 1 1 2 1 & & $(-0.2519,0.0686)$& 0.068\\   
           & &4 2 0 2 0 & & $(0.1672,0.0007)$& 0.028\\    
           & &4 1 3 2 1 & & $(0.1036,-0.0025)$& 0.011\\   
           & &4 3 1 2 1 & & $(0.1029,0.0064)$& 0.011\\    
           & &4 1 1 1 1 & & $(0.1223,-0.0123)$& 0.015\\   
     
     \botrule
    \end{tabular}
    \caption{Asymptotic coefficients $A'_\beta$ and their squared moduli for the 0$^+$ and 2$^+$ resonances. Only the most representative channels $\beta\equiv\{K,l_x,l_y,l,S_x\}$, with significant weights, are shown in each case. A global phase has been added to make the first channel real.}
    \label{tab:coef}
\end{table}

\subsection{Relative-energy distributions}
\begin{figure}
\centering
\includegraphics[width=0.85\linewidth]{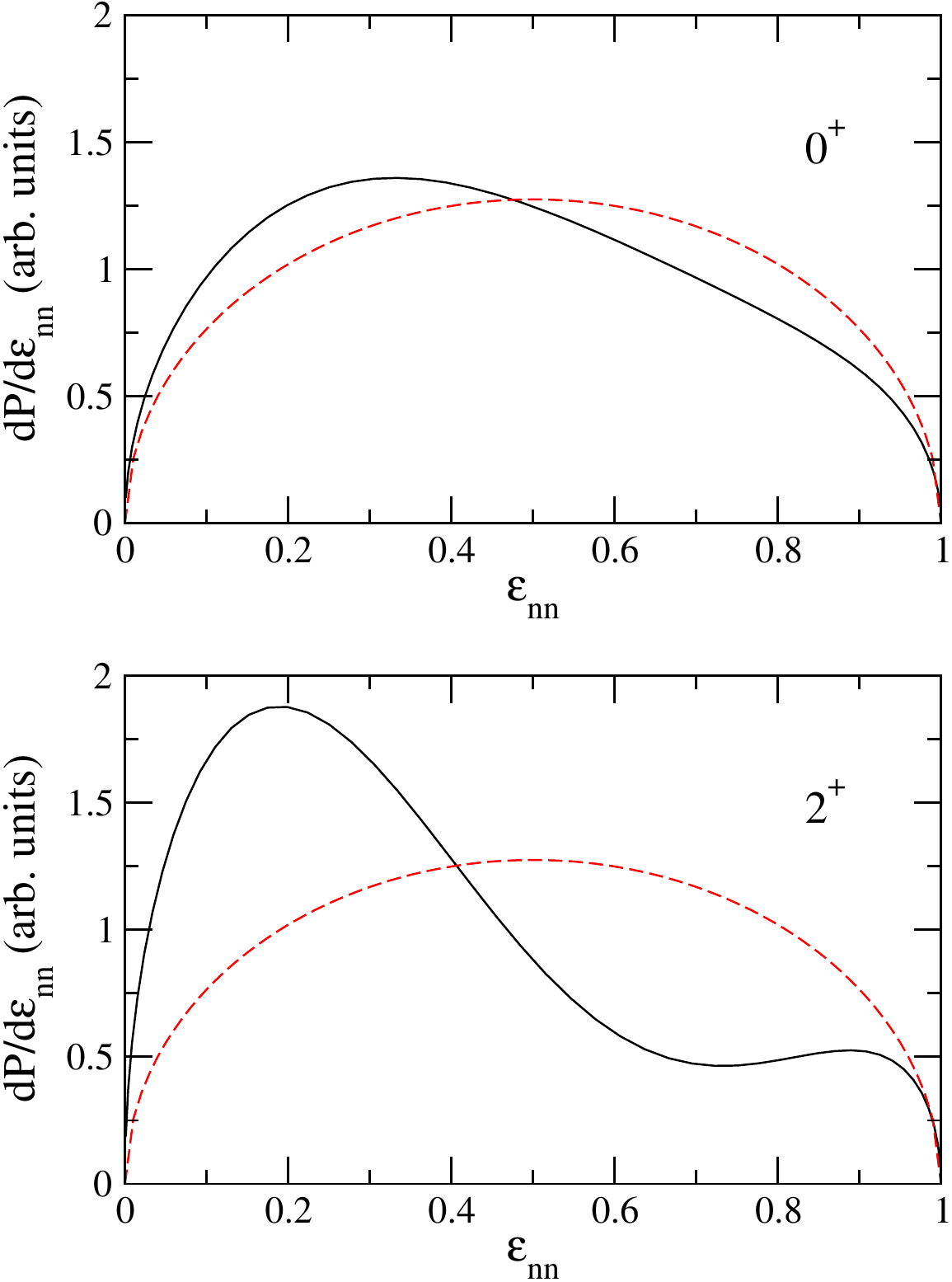}

\caption{Neutron-neutron relative energy distributions for the 0$^+$ (top panel) and 2$^+$ (bottom panel) resonances. The dashed line corresponds to three-body phase-space decay. Distributions are normalized to unity.}
\label{fig:enn}
\end{figure}

Once the resonance wave functions at large distances have been determined, the corresponding asymptotic coefficients describe the energy correlations in Eqs.~(\ref{eq:Palpha}-\ref{eq:Pex}). The  normalized $n$-$n$ relative-energy distribution (including all asymptotic channels), for both the 0$^+$ and 2$^+$ resonances, is shown in Fig.~\ref{fig:enn}. For comparison, the lines corresponding to phase-space decay (no correlations)~\cite{grigorenko18}, $\frac{8}{\pi}\sqrt{\varepsilon(1-\varepsilon)}$, is also shown. 

For the $0^+$ state, the dominant $K=0$ term, with a single channel $\beta=\{0,0,0,0,0\}$, induces a distribution that follows the phase space. Then, its coherent sum with the $K=2$ channel $\beta=\{2,0,0,0,0\}$ produces a deviation from the phase-space curve and a maximum at around $\varepsilon_{nn}\approx 0.3$. The position of this maximum arises from the relative phase between the two components, which, in this case, favors low $\varepsilon_{nn}$ values. 

\begin{figure}
\centering
\includegraphics[width=0.85\linewidth]{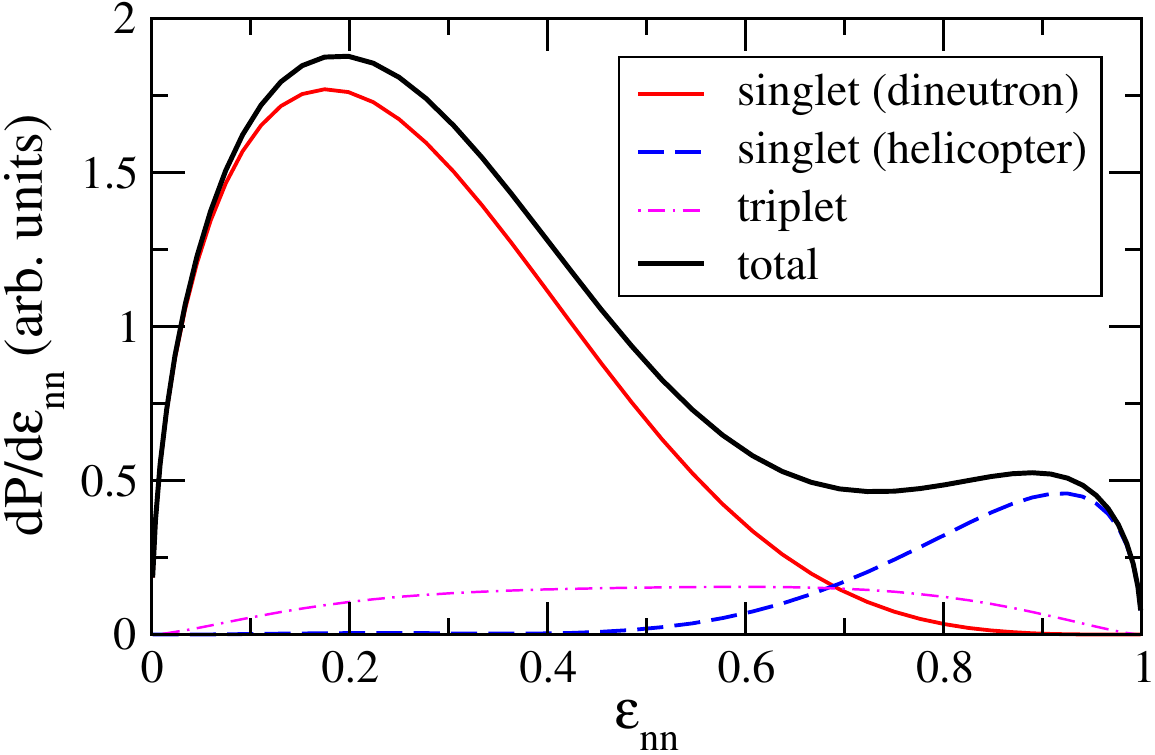}

\caption{Contributions to the neutron-neutron relative-energy distribution of the 2$^+$ state. The singlet dineutron ($K=2, l_x=0,S_x=0$, red solid line), helicopter ($l_x=2,S_x=0$, dashed blue) and triplet ($S_x=1$, dot-dashed magenta) components are shown. See the text.}
\label{fig:2pdineutron}
\end{figure}

For the $2^+$ state, since the $K=0$ term is not allowed, the distribution reflects the mixing between $K=2$ and $K=4$ components. The result is a more pronounced maximum at around $\varepsilon_{nn}\approx 0.2$ and a two-peak structure. To understand this result, the three contributions shown in Fig.~\ref{fig:2pdineutron} can be identified. First, there is a large contribution arising from the dominant channel $\beta=\{2,0,2,2,0\}$, which generates a broad distribution at low $\varepsilon_{nn}$ values peaked at $\approx0.2$. Note that this channel could mix coherently with all channels defined by $l_x=0,l_y=2,l=2$ but different $K$, however their norm is negligible. In that sense, we may call this a ``dineutron'' contribution ($lx=0, S_x=0$), coming mainly from a single asymptotic channel. Then there is a relevant contribution arising from $\beta=\{2,2,0,2,0\}$, interfering with the component $\beta=\{4,0,2,2,0\}$, which gives rise to a sharp (but small) peak at large $\varepsilon_{nn}\approx0.9$. This can be visualized as a ``helicopter'' component ($l_x=2, S_x=0$), where the neutrons have large relative angular momentum and large relative energy. Note that both the dineutron and helicopter correspond to the neutron-neutron spin singlet. Finally, there are several spin triplet components, characterized by $S_x=1$, and dominated by $\beta=\{2,1,1,1,1\}$, which produce a broad continuum dominated by energies around $\varepsilon_{nn}= 0.5$. 

Interestingly, the relative-energy spectrum for the 2$^+$ resonance is more peaked at low $n$-$n$ energies than that of the 0$^+$ ground state, although the spatial distribution of the source states (see Fig.~\ref{fig:xyprob}) is less compact for the 2$^+$. As discussed in Ref.~\cite{Monteagudo2024}, a simple FSI model (such as Ref.~\cite{lednicky1982}) would lead to the conclusion that a more pronounced low-$\varepsilon_{nn}$ peak comes from an overall smaller neutron source. This highlights the need to employ proper three-body decay models that can account for dynamical effects, in particular the wave-function evolution and effective barriers discussed above. The qualitative agreement between the theoretical results and the experimental data, already discussed in Ref.~\cite{Monteagudo2024}, supports this fact. 

\begin{figure}
\centering
\includegraphics[width=0.85\linewidth]{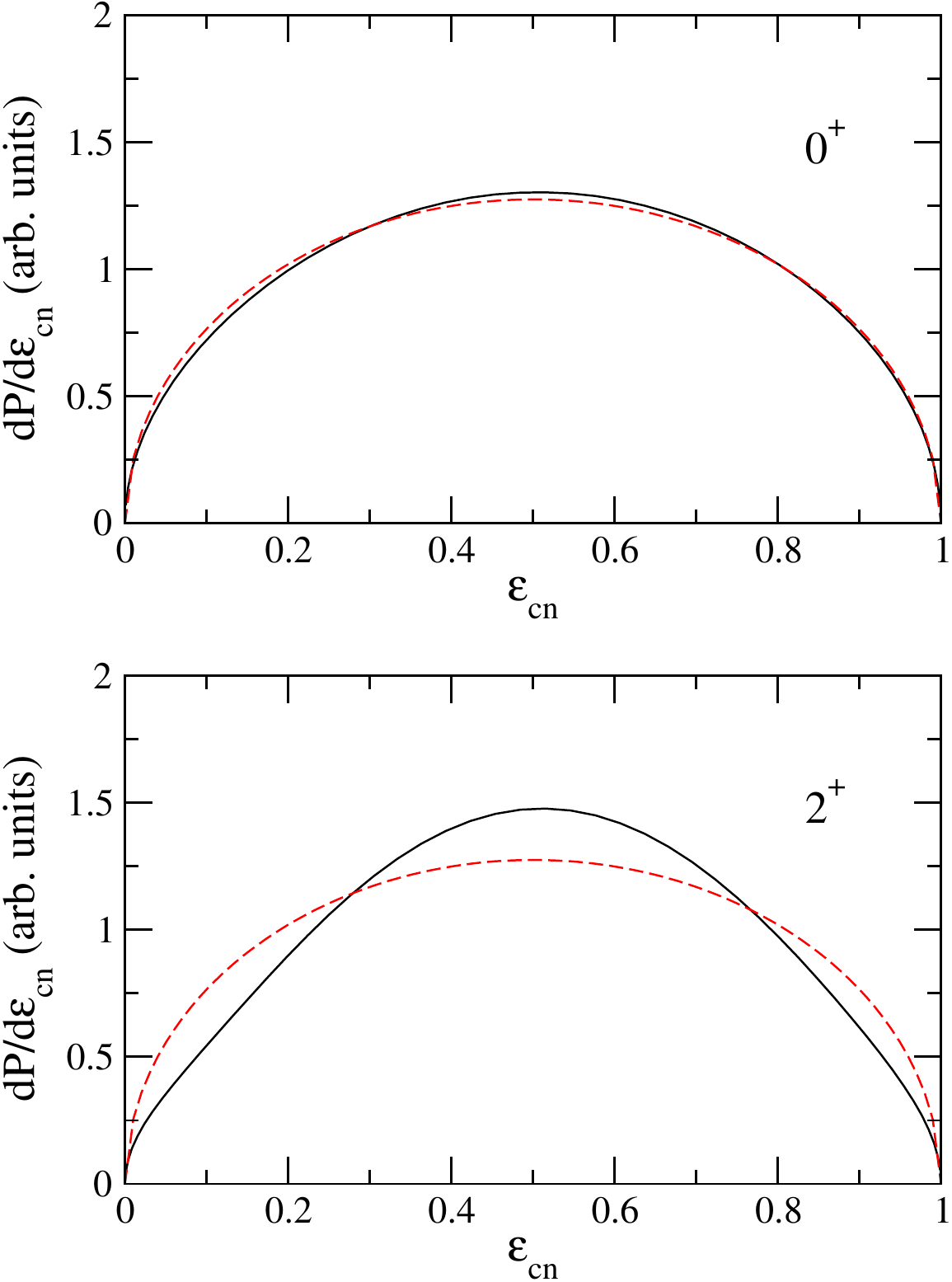}

\caption{The same as Fig.~\ref{fig:enn} but for the core-neutron relative energy.}
\label{fig:ecn}
\end{figure}

By transforming the asymptotic solution from the Jacobi-T to the Jacobi-Y representation, where the core and one neutron are related by the $x$ coordinate, the $\text{core}+n$ relative-energy distribution can be calculated. This is shown in Fig.~\ref{fig:ecn}. The result for the 0$^+$ resonance is quite similar to the phase-space line, while that for the 2$^+$ presents a somewhat symmetric but narrower distribution. This indicates direct decay of the two resonances, and is consistent with the discussion in Ref.~\cite{Monteagudo2024} from the absence of peaks associated to resonances in the intermediate $^{15}$Be ($^{14}\text{Be}+n$) system.

\section{Conclusions}

We have developed and applied a general formalism to describe the decay of three-body resonances, using an inhomogeneous Schrödinger equation and a discrete pseudostate representation. By formally  separating the short- and long-range components of the resonance wave function, and by identifying physically meaningful short-range source states from the eigenstates of a resonance operator, the method provides a clean and computationally efficient way to access the asymptotic properties of the decaying system. The resulting outgoing-wave solution, constructed from the free propagator and the modified source term, allows us to extract relative-energy distributions without requiring 
interaction-dependent continuum wave functions or 
complex-energy eigenstates. 

The formalism has been applied to the unbound nucleus $^{16}$Be ($^{14}\text{Be}+n+n$), demonstrating the capability of the method to capture the essential correlations that govern direct two-neutron emission. Using the hyperspherical expansion method, realistic effective interactions, and a large THO basis, we have studied the 0$^+$ ground-state and 2$^+$ excited-state resonances, analyzing their internal structure, source terms, and asymptotic amplitudes. For both resonances, the asymptotic behavior is dominated by the lowest hypermomentum components, illustrating how the formalism naturally incorporates barrier-penetration effects and the dynamical evolution of the resonant source state as it decays. 
The computed neutron–neutron and core–neutron relative-energy spectra exhibit characteristic signatures of direct three-body decay, and display qualitative agreement with recent experimental measurements. In particular, the enhanced low-energy $nn$ correlations obtained for the 2$^+$ resonance (despite its less compact initial configuration) highlight the importance of dynamical effects beyond simple FSI models. This reinforces the view that correct asymptotic three-body dynamics are essential for a meaningful interpretation of multi-neutron decay observables. 

In summary, the formalism presented here provides a robust framework for connecting short-range few-body structure with measurable decay correlations in systems beyond the dripline. Extensions of this work include the possibility of incorporating core excitations, refined two-body and three-body interactions, as well as the calculation of angular distributions in addition to energy distributions. The method could be applied to other two-neutron emitters and to describe the decay of three-body resonances in Borromean nuclei.

\section*{Acknowledgments}
We thank B.~Monteagudo, F.~M.~Marqués and collaborators for discussions about the experimental results in Ref.~\cite{Monteagudo2024} and for allowing us to contribute to the theoretical interpretation. This work has been funded by MICIU/AEI/10.13039/501100011033 
and ERDF/EU [projects No.~PID2023-146401NB-I00, PID2021-123879OB-C21],  by Horizon 2020 under the Marie Sk{\l}odowska-Curie Actions [Grant Agreement No.~101023609] and by Horizon Europe under the Research Infrastructures programme [Grant Agreement No.~101057511].

\section*{Appendix}
Let us state the criterion to determine the source state.  A resonance involves a significant increase in cross sections, and this will occur provided that the norm of the modified source term including interactions is large. Thus, from all possible normalized states in the short range subspace, we will consider the source state $|\phi_r \rangle$ that maximizes the norm of the modified source term, which is given by the positive-definite number 
  \begin{eqnarray}
  \langle \Phi_s   |\Phi_s \rangle     &=&  \sum_{n m m'}  \langle \phi_r |m' \rangle    \langle m |\phi_r \rangle    \nonumber \\
   && \times\left({ \langle m' | V | n \rangle \over \epsilon_{m'} -E^*_r} - \delta_{nm'} \right) \nonumber \\ && \times\left({ \langle n | V | m \rangle \over \epsilon_m -E_r} - \delta_{nm} \right).
  \end{eqnarray}  
The states which make this norm stationary will be precisely the eigenstates of the operator 
\begin{eqnarray}
    O^2 &=& \left({1 \over H_{ef} -E^*_r} V - 1 \right) \left( V {1 \over H_{ef} -E_r}- 1 \right)  \label{OperatorO} \nonumber \\
    &=& {1 \over H_{ef} -E^*_r} (T - E^*_r) (T - E_r) {1 \over H_{ef} -E_r}.
\end{eqnarray}
In particular, the sharpest resonance, which will likely produce the most pronounced increase in the cross section, is linked to the largest possible eigenvalue of $O^2$.

For the source term including interaction and energy width , we have
  \begin{eqnarray}
  \langle {\Phi'}_s   |{\Phi'}_s \rangle     &=&  \sum_{n m m'}  \langle \phi_r |m' \rangle    \langle m |\phi_r \rangle    \nonumber \\
   && \times \left({ \langle m' | V | n \rangle - i {\Gamma\over 2 } \delta_{m'n} \over \epsilon_{m'} -E^*_r} - \delta_{nm'} \right) \nonumber \\ && \times \left({ \langle n | V | m \rangle  + i {\Gamma\over 2 } \delta_{mn}   \over \epsilon_m -E_r} - \delta_{nm} \right) .
  \end{eqnarray}  
In this case, the states which make this norm stationary will be the eigenstates of
\begin{eqnarray}
    \widetilde{O}^2 &=& \left({1 \over H_{ef} -E^*_r} (V -  i {\Gamma\over 2 }) - 1 \right) \nonumber \\
    &\times & \left( (V +  i {\Gamma\over 2 }) {1 \over H_{ef} -E_r}- 1 \right) \label{OperatorOtilde} \nonumber \\
      &=& {1 \over H_{ef} -E^*_r} (T - e_r) (T - e_r) {1 \over H_{ef} -E_r}.
\end{eqnarray}

We can make some approximations in these expression, in order to connect this treatment with our previous, heuristic definition of the resonance in Ref.~\cite{JCasal19}. First, let us neglect the resonance energy $E_r$ in comparison with the typical eigenvalues of the Hamiltonian in the discrete basis.   Consistently with this, ${\Gamma \over 2} $ is negligible compared to the interaction. Then, we get a Hermitian, positive-definite operator $\widetilde{M}^2$, which is independent on the energy of the resonance, given by
\begin{equation}
\widetilde{M}^2 =  \left({1 \over H_{ef}} V - 1 \right) \left( V {1 \over H_{ef}}- 1 \right) \simeq  \widetilde{O}^2 \simeq  O^2.
\end{equation}
Now, we consider situations in which the relevant Hilbert space only contains states with positive energy.  Thus, bound states are excluded, or projected out through the operator $P$. In this case, we can define the Hermitian operator $H_{ef}^{-1/2}$. 
Then we can assume that the commutator of the Hamiltonian with the interaction $[H,V]$ is small compared to the corresponding anticommutator $\{H,V\}$. This allows us to symmetrize the previous expression, to give 
\begin{equation}
 \widetilde{M}^2 \simeq  (M-1)^2 = \left(H_{ef}^{-1/2} V H_{ef}^{-1/2}- 1 \right)^2 .
\end{equation}
Indeed, this operator commutes with the resonance operator $M = H_{ef}^{-1/2} V  H_{ef}^{-1/2}$ used in our previous work~\cite{JCasal19}. Eigenstates of $M$ that have large (negative) eigenvalues  should be close to the eigenstates $  \widetilde{M}^2 $,  and assuming that the resonance energies are small, also to the eigenstates of $O^2, \widetilde{O}^2$ which give the maximum norm of the modified source term. In this way, we find here a justification for our previous selection of the resonance. 

\bibliography{newref}

\end{document}